\documentclass{article}

\usepackage{arxiv}

\usepackage[utf8]{inputenc} 
\usepackage[T1]{fontenc}    
\usepackage{url}            
\usepackage{booktabs}       
\usepackage{amsfonts}       
\usepackage{nicefrac}       
\usepackage{microtype}      
\usepackage{graphicx}
\usepackage[numbers, square]{natbib}
\usepackage{doi}

\usepackage{authblk}
\usepackage{gensymb}
\usepackage{chemformula}
\setchemformula{kroeger-vink}
\title{Optical and microstructural characterization of \\ Er$^{3+}$ doped epitaxial cerium oxide on silicon}

\author[1,2]{\large Gregory D. Grant}
\author[1,3]{Jiefei Zhang}
\author[1,2]{Ignas Masiulionis}
\author[1]{Swarnabha Chattaraj}
\author[1]{Kathryn E. Sautter}
\author[1]{Sean E. Sullivan\thanks{Present address: memQ, Inc. Chicago, Illinois 60615 , United States.} \hspace{1px}}
\author[2]{Rishi Chebrolu}
\author[4]{Yuzi Liu}
\author[5]{Jessica B. Martins}
\author[6]{Jens Niklas}
\author[1,3,7]{Alan M. Dibos}
\author[8]{Sumit Kewalramani}
\author[5]{John W. Freeland}
\author[4]{Jianguo Wen}
\author[6]{Oleg G. Poluektov}
\author[1,2,3]{F. Joseph Heremans}
\author[1,2,3]{David D. Awschalom}
\author[1,2,3]{Supratik Guha\thanks{Corresponding author.  Contact at sguha@anl.gov} \hspace{1px}}

\affil[1]{\normalsize Materials Science Division, Argonne National Laboratory, Lemont, Illinois 60439, United States}
\affil[2]{Pritzker School of Molecular Engineering, University of Chicago, Chicago, Illinois 60637, United States}
\affil[3]{Center for Molecular Engineering, Argonne National Laboratory, Lemont, Illinois 60439, United States}
\affil[4]{Center for Nanoscale Materials, Argonne National Laboratory, Lemont, Illinois 60439, United States}
\affil[5]{X-ray Science Division, Argonne National Laboratory, Lemont, Illinois 60439, United States}
\affil[6]{Chemical Sciences and Engineering Division, Argonne National Laboratory, Lemont, Illinois 60439, United States}
\affil[7]{Nanoscience and Technology Division, Argonne National Laboratory, Lemont, Illinois 60439, United States}
\affil[8]{Department of Materials Science and Engineering, Northwestern University, Evanston, Illinois 60208, United States}

\date{}

\renewcommand{\headeright}{}
\renewcommand{\undertitle}{}
\renewcommand{\shorttitle}{Optical and microstructural characterization of Er$^{3+}$ doped epitaxial cerium oxide on silicon}

\begin{document}
\maketitle

\begin{abstract}
Rare-earth ion dopants in solid-state hosts are ideal candidates for quantum communication technologies such as quantum memory, due to the intrinsic spin-photon interface of the rare-earth ion combined with the integration methods available in the solid-state.  Erbium-doped cerium oxide (\ch{Er:CeO2}) is a particularly promising platform for such a quantum memory, as it combines the telecom-wavelength ($\sim1.5$ $\mu$m) 4f-4f transition of erbium, a predicted long electron spin coherence time supported by \ch{CeO2}, and is also near lattice-matched to silicon for heteroepitaxial growth.  In this work, we report on the epitaxial growth of \ch{Er:CeO2} thin films on silicon using molecular beam epitaxy (MBE), with controlled erbium concentration down to 2 parts per million (ppm).  We carry out a detailed microstructural study to verify the \ch{CeO2} host structure, and characterize the spin and optical properties of the embedded \ch{Er^{3+}} ions.  In the 2-3 ppm Er regime, we identify EPR linewidths of 245(1) MHz, optical inhomogeneous linewidths of 9.5(2) GHz, optical excited state lifetimes of 3.5(1) ms, and spectral diffusion-limited homogenoeus linewidths as narrow as 4.8(3) MHz in the as-grown material.  We test annealing of the \ch{Er:CeO2} films up to 900 \celsius{}, which yields modest narrowing of the inhomogeneous linewidth by 20\%{} and extension of the excited state lifetime by 40\%{}. We have also studied the variation of the optical properties as a function of Er doping and find that the results are consistent with the trends expected from inter-dopant charge interactions.
\end{abstract}


\section{Introduction}

Erbium (Er) doped solid-state hosts present a quality candidate for quantum memory \cite{awschalom2022roadmap, awschalom2021development, Kinos2021}, due to the intrinsic spin-photon interface and long coherence times of Er \cite{liu2006spectroscopic}. Additionally, the compatibility of Er with the telecom C-band \cite{Wolfowicz2021} due to its 1.53 $\mu$m optical transition and the versatility of solid-state hosts \cite{Wolfowicz2021} has motivated a multitude of recent efforts in different erbium-doped host materials such as \ch{Er:TiO2} \cite{singh2022development, dibos2022purcell, ji2023nanocavitymediated}, \ch{Er:Y2O3} \cite{singh2020epitaxial, gupta2023towards}, \ch{Er:CaWO4} \cite{le2021twenty, Ourari2023}, \ch{Er:Y2SiO5} \cite{ranvcic2018coherence}, \ch{Er:MgO} \cite{horvath2023strong}, \ch{Er:YVO4} \cite{rochman2023microwave}, \ch{Er:LiNbO3} \cite{yu2023frequency}, and \ch{Er:Si} \cite{Weiss21, rinner2023erbium} in order to explore various desired characteristics.

Optimizing the electron spin coherence of \ch{Er^{3+}} for quantum memory applications necessitates selection of an environment free from decoherence mechanisms, and for high-quality wide-bandgap crystals at cryogenic temperatures the leading factor of decoherence is nearby nuclear spins within the host material \cite{kanai2022generalized}.  A recent computational study identified that cerium dioxide (\ch{CeO2}) is an optimal host for maximizing electron spin coherence \cite{kanai2022generalized}, due to the near-zero natural abundance of nuclear spins in its constituent elements \cite{stoll2006easyspin}.  \ch{CeO2} is additionally attractive as a host due to its low lattice mismatch ($-0.4$\%{}) with silicon (Si), with heteroepitaxy on silicon providing an avenue for scalability of photonic and electronic quantum devices, as discussed in more detail elsewhere \cite{dibos2022purcell, singh2020epitaxial}.

In this work, we benchmark Er-doped \ch{CeO2} (\ch{Er:CeO2}) thin films grown on Si(111) by molecular beam epitaxy (MBE) for use in developing a telecom-wavelength interfaced spin qubit platform.  Previous work on \ch{CeO2}/Si epitaxy has focused mostly on microstructural studies \cite{yoshimoto1990situ, nagata1992type, nishikawa2002electrical, barth2016perfectly}.  Inaba \textit{et al.} \cite{inaba2018epitaxial} have examined \ch{Er:CeO2}/Si down to 10,000 parts per million (ppm) Er and reported preliminary \ch{Er^{3+}} optical characterization results at 4 K, including an optical excited state lifetime ($T_1$) of 1.5 ms at 1512 nm.  We extend this line of investigation by exploring significantly lower Er concentration regimes (1-100 ppm) than have been studied previously for \ch{CeO2}.

We conduct a detailed study of the MBE-grown \ch{Er:CeO2}/Si(111) system, starting with microstructural study of the thin film where we confirm that \ch{CeO2} grows epitaxially on the Si(111) substrate with appropriate \ch{Ce^{4+}} valency.  Examining the spin properties of the Er doped into the \ch{CeO2} system by electron paramagnetic resonance (EPR), we identify results consistent with \ch{Er^{3+}} substituting into the cubic \ch{Ce^{4+}} site, and at concentrations of 2-3 ppm we measure EPR linewidths as low as $245(1)$ MHz.

We additionally examine the optical properties of the trivalent \ch{Er^{3+}} dopants at 3.5 K, identifying an optical inhomogeneous linewidth of $9.5(2)$ GHz, a spectral diffusion linewidth as narrow as $4.8(3)$ MHz, and an optical lifetime as long as $3.5(1)$ ms at 2-3 ppm doping levels.  Of particular note is that the spectral diffusion linewidths found here via transient spectral hole burning -- though broader than in bulk or with nanostructures built upon bulk samples as measured by spectral hole burning \cite{tawara2017effect} or photon echo \cite{Ourari2023, thiel2011rare, wang2022er} -- are narrower by an order of magnitude than other reports of spectral diffusion in thin film or nanostructured Er-doped oxides on silicon \cite{singh2022development, dibos2022purcell, ji2023nanocavitymediated}.

To elaborate upon our study we identify the trends by which the EPR, optical inhomogeneous, and spectral diffusion linewidths narrow and the excited state lifetime increases as a function of decreasing Er concentration.  Additionally, we examine the effect of annealing the \ch{Er:CeO2} films up to 900 \celsius{}, which yields modest narrowing of the inhomogeneous linewidth by 20\%{} and extension of the excited state lifetime by 40\%{}.

\section{Methods}

\subsection{Epitaxial growth of \ch{Er:CeO2}}

\ch{Er:CeO2} thin films are grown epitaxially on Si(111)$\pm0.5\degree$ substrates using a Riber C21 DZ Cluster molecular beam epitaxy (MBE) system. Growths are carried out between 665-675 \celsius{} and initiated on a $7\times7$ reconstructed Si(111) surface. Metallic Er, Ce and molecular \ch{O2} beams are used with a beam equivalent \ch{O2}:\ch{Ce} flux ratio of 20.  Growths are observed \textit{in situ} with a reflection high-energy electron diffraction (RHEED) system operated at 15 kV.  We grow with erbium doping levels between 2-132 ppm and thicknesses between 200-940 ppm.  Following deposition, films are cooled in the presence of oxygen flux. Further details are described in the supplemental information (S.I.). Where anneals are noted, samples are annealed after growth in an MTI OTF-1200X tube furnace at one atmosphere of 20.02\%{} \ch{O2} balanced with \ch{Ar}.

\subsection{Structural, spin, and optical characterization of epitaxial films}

Cross-sectional XTEM studies are carried using a Thermo Fisher Spectra 200 operated at 200 keV in scanning transmission electron microscopy (STEM) mode.  The same tool at the same voltage is used for energy-dispersive x-ray (EDX) spectroscopy. Specimens for XTEM and EDX are prepared using focused ion beam (FIB) milling \cite{mayer2007tem}.  A four-circle Rigaku Smartlab diffractometer is used for X-ray diffration (XRD) scans, along with X-ray absorption spectroscopy (XAS) performed on beamline 29 ID-D of the Advanced Photon Source at Argonne National Laboratory.  All microstructural measurements were conducted at room temperature.

Continuous wave (CW) X-band (9.5 GHz) EPR experiments are carried out with a Bruker ELEXSYS II E500 EPR spectrometer (Bruker BioSpin), equipped with a TE$_{102}$ rectangular EPR resonator (Bruker ER 4102ST).  Field modulation at 100 kHz in combination with lock-in detection leads to first derivative-type CW EPR spectra.  Measurements are performed at cryogenic temperatures between 4.0 and 4.2 K, with temperature governed by a helium gas-flow cryostat (ICE Oxford) and an ITC (Oxford Instruments).  \ch{Er:CeO2} samples are mounted with the static magnetic field parallel to the Si$<110>$ axis. Measurements use a field modulation of 1 mT, a microwave power attenuation of 35 dB (from 200 mW), and a field step size of 0.2 mT.

Optical characterization is performed in a custom confocal microscopy setup designed for telecom C-band spectroscopy, with samples mounted in a cryostat at 3.5 K (s50 Cryostation, Montana Instruments).  Time-resolved photoluminescence excitation (PLE) spectroscopy is done with 1.5 ms excitation pulses shaped by acousto-optic modulators (AOMs) and 7 ms collection intervals. The PLE signal is detected by a Quantum Opus superconducting nanowire single photon detector (SNSPD).  Transient spectral hole burning (TSHB) measurements are enabled by the addition of a phase electro-optic modulator to the excitation path, yielding sidebands with a specific detuning from the laser carrier.  Photoluminescence (PL) measurements are performed in the same setup with an alternative collection path routed to a low-noise InGaAs camera (PyLoN IR, Princeton Instruments) and the excitation laser operated continuously.  Additional details of this setup are described elsewhere \cite{ji2023nanocavitymediated}.

\section{Results and Discussion}

\begin{figure}[t]
  \centering
  \includegraphics[width=0.75\linewidth]{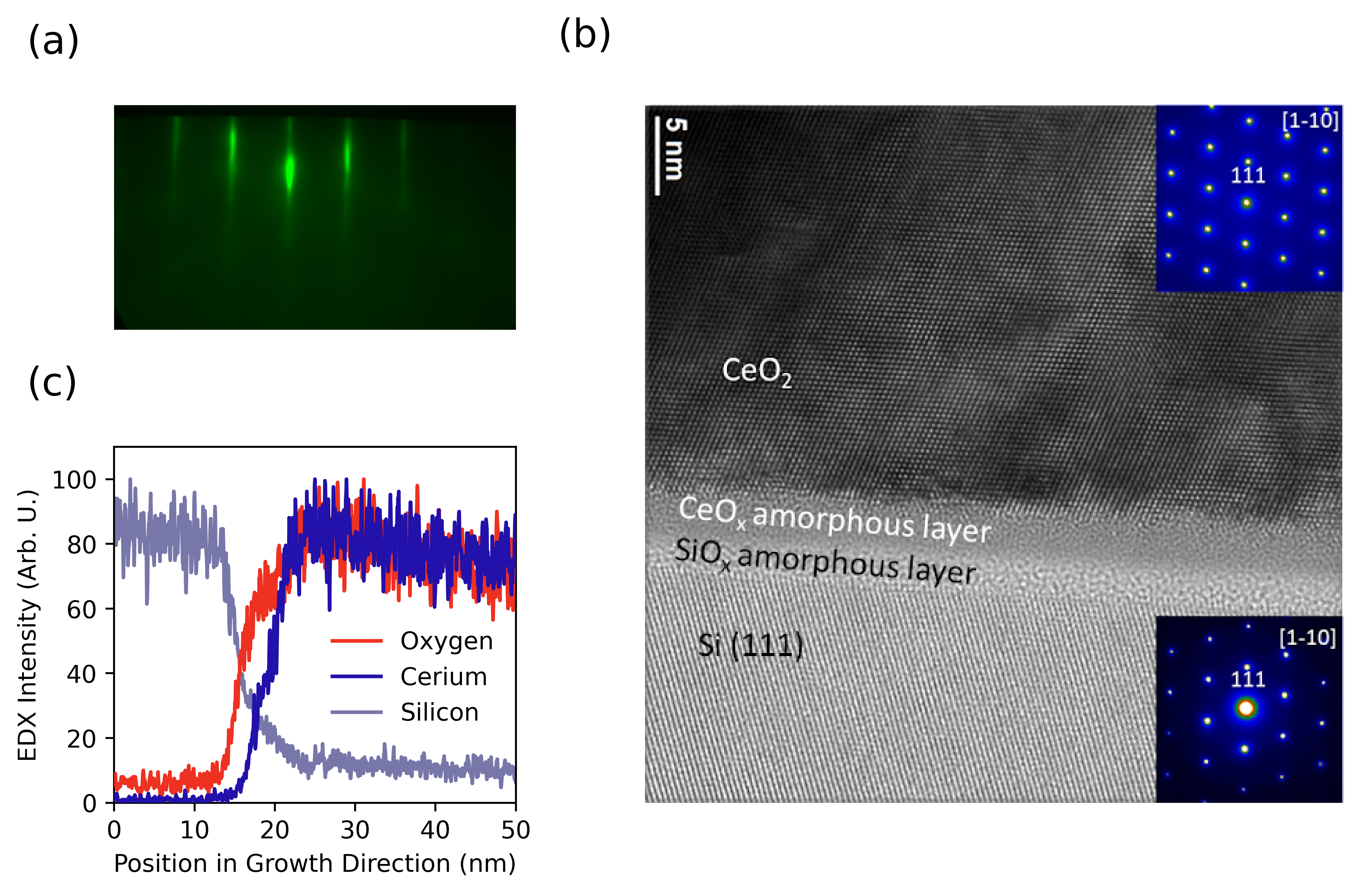}
  \caption{Epitaxy of a 3 ppm Er, 940 nm thick \ch{CeO2} thin film on Si. (a) RHEED pattern of the epitaxial \ch{CeO2} surface with the electron beam along the Si$<110>$ azimuthal direction. (b) High resolution cross-sectional TEM of the \ch{Er:CeO2}/Si structure showing epitaxial registry of the \ch{CeO2} with the Si substrate, as well as a 4 nm thick amorphous layer between the film and substrate.  Diffraction patterns are shown in the insets.  (c) EDX scan across the \ch{CeO2}/Si interface, showing that the amorphous layer is composed of a mixed \ch{Ce} and Si oxide. Each intensity trace is normalized to the maximum counts for that element.}
  \label{fig:fig1}
  \rule{\textwidth}{1pt}
\end{figure}

\subsection{Microstructural characterization of \ch{Er:CeO2}}

Figure~\ref{fig:fig1}(a) shows the RHEED pattern of a 3 ppm Er, 940 nm thick \ch{CeO2} film on Si(111) immediately following growth, with the electron beam incident along the Si$<110>$ azimuth. No significant changes in RHEED patterns are noted during growth or for different Er doping concentrations.  The streaky pattern indicates a smooth, single-crystalline surface and is consistent with an epitaxial \ch{CeO2}(111)/Si(111) alignment between the epilayer and substrate. This is further confirmed by cross-sectional TEM studies. Figure~\ref{fig:fig1}(b) shows a representative high-resolution bright field XTEM image with diffraction patterns shown in the inset.  A 4 nm thick amorphous layer is observed at the \ch{CeO2}/Si interface.  We identified a mixed \ch{CeO}$_x$-\ch{SiO}$_y$ composition across this layer via EDX (Figure~\ref{fig:fig1}(c)). Similar interfacial oxide layers have been observed in \ch{CeO2}/Si previously \cite{inaba2018epitaxial} and are a well-known phenomenon in ionic oxides grown on Si \cite{singh2020epitaxial, narayanan2002interfacial}, resulting from oxygen diffusion followed by catalytic oxidation of the buried silicon interface.

Low-magnification bright field XTEM (see S.I. Figure~\ref{si:fig1} for an example) shows an epilayer threading dislocation density of $\sim10^9$ cm$^{-2}$.  Similar threading defects can be seen in the XTEM studies of \ch{CeO2}/Si by Inaba \textit{et al.} \cite{inaba2018epitaxial}.  Threading segments do not relieve lattice mismatch strain, and we ascribe the formation of these threading defects to the initial stages of epitaxial growth of \ch{CeO2}, possibly due to the formation of localized patches of oxidized silicon due to catalytic effects of the deposited \ch{CeO2}. This effect may be controlled by adjusting the growth conditions and will be the subject of a later paper.

\begin{figure}[t]
  \centering
  \includegraphics[width=0.75\linewidth]{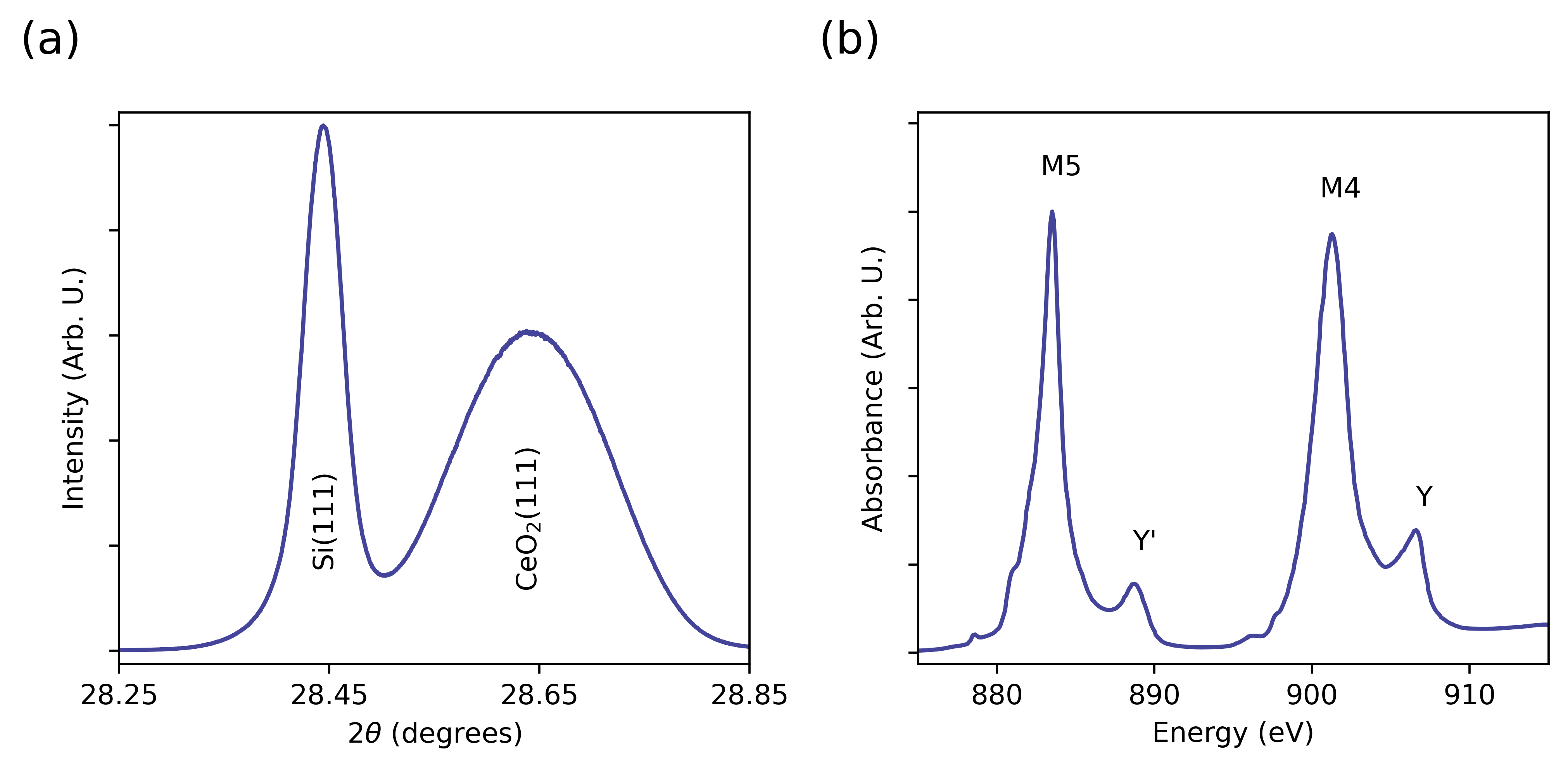}
  \caption{Additional microstructural study of as-grown \ch{Er:CeO2} samples.  (a) $\omega$-$2\theta$ XRD scan of a 3 ppm Er, 940 nm thick \ch{CeO2} thin film on Si.  The FWHM of the \ch{CeO2} peak is 630 arcsec. (b) X-ray absorption spectroscopy of a 35 ppm Er, 240 nm thick \ch{CeO2} thin film on Si.  The cerium M-edge is shown, as detected by total electron yield (TEY) mode and normalized to the maximum measured intensity. Consistent with \ch{Ce^{4+}}, the M5 and M4 peaks are at $883$ eV and $901$ eV respectively.  The satellite peaks Y' and Y are observed at $889$ eV and $906.5$ eV respectively.}
  \label{fig:fig2}
  \rule{\textwidth}{1pt}
\end{figure}

An $\omega$-$2\theta$ XRD scan of a 3 ppm Er, 940 nm thick \ch{CeO2} sample, as shown in Figure~\ref{fig:fig2}(a), yields a \ch{CeO2}(111) refection with a  full-width at half maximum (FWHM) of  630 arcsec, qualitatively consistent with the threading dislocation density observed. The peak separation between the Si(111) and \ch{CeO2}(111) peaks is $\sim$700 arcsec indicating that most of the misfit strain remains elastically stored in the film.

To corroborate the crystal structure identified by XRD, XAS of the \ch{Ce} M-edge on a 35 ppm Er, 240 nm thick \ch{CeO2} on Si sample shows two sets of peaks related to the M5 and M4 transitions of electrons from 3d core orbitals to unoccupied p- and f-like symmetry orbitals, as seen in Figure~\ref{fig:fig2}(b). The positions of the main peaks at 883 eV (M5) and 901 eV (M4) relate to the electric-dipole allowed transitions to 4f states \cite{thole19853d, paidi2019role, chen2012concentration} and are consistent with the \ch{Ce^{4+}} valence state and the formation of \ch{CeO2} (as opposed to \ch{Ce^{3+}} and \ch{Ce2O3}). The satellite peaks at 889 eV (Y') and 906.5 (Y) result from transition to 4f states in the condition band and are additional indicators of predominately \ch{Ce^{4+}} valency \cite{garvie1999ce4ce3, hayakawa2016x}.  Overall, spectral shape and peak separation are consistent with the \ch{Ce^{4+}} oxidation state, and peaks corresponding to \ch{Ce^{3+}} are not identifiable within the spectrum.  This data together with the XRD and XTEM studies suggests that we have a \ch{CeO2} film where the \ch{Ce^{3+}} concentration is likely less than 1\%{}, based on the detection limit of the experimental setup.

\subsection{EPR study on erbium incorporation into \ch{CeO2}}

\begin{figure}[t]
  \vspace{0.5cm}
  \centering
  \includegraphics[width=0.75\linewidth]{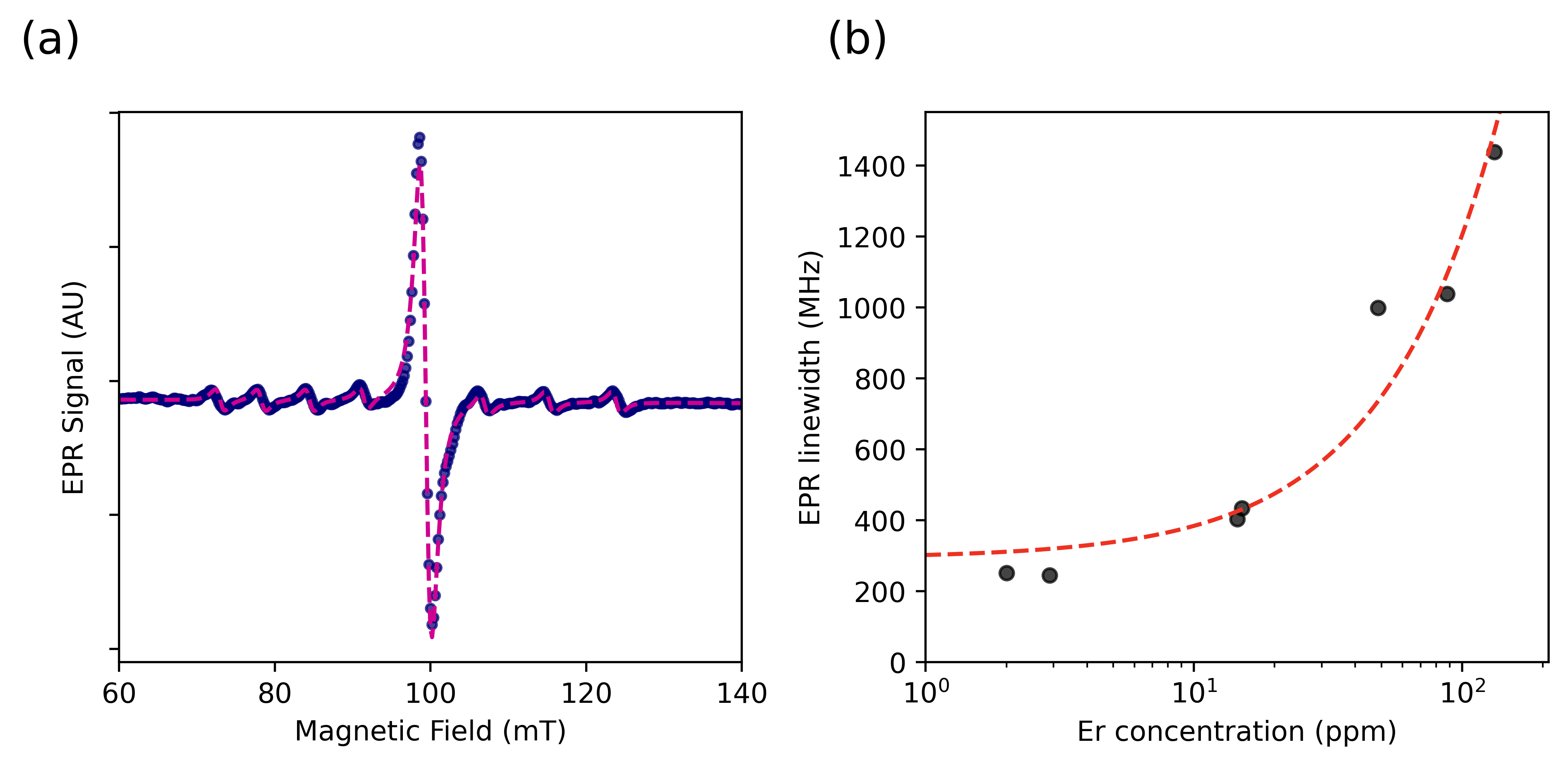}
  \caption{EPR study of \ch{Er:CeO2} thin films on Si.  EPR measurements are performed at 4.0-4.2 K. (a) CW EPR resonance spectrum of \ch{Er:CeO2}, obtained from a 3 ppm Er, 940 nm thick sample.  A primary peak at $\sim 100$ mT is produced by nuclear spin zero Er isotopes, and secondary peaks due to the less abundant \ch{^{167}Er} are visible around the main peak.  The peak locations are obtained via fit using Eq.\ref{zeeman} (magenta dashed line), and we obtain a $g$-value of $g = 6.812(5)$ and a hyperfine splitting parameter of $A=687(1)$ MHz.  (b)  The Er spin resonance linewidth as a function of Er concentration, extracted from the primary peak of the CW EPR spectrum at each Er concentration (black circles).  Uncertainty in the extracted linewidths are smaller than the data marker size.  A linear fit to the Er concentration (red dashed line) matches the trend of the data, and is discussed further in the main text.}
  \label{fig:fig3}
  \vspace{0.5cm}
  \rule{\textwidth}{1pt}
\end{figure}

The incorporation of \ch{Er^{3+}} into the \ch{CeO2} films is confirmed by identifying erbium-specific spin properties under EPR.  Low-temperature EPR probes the lowest-lying level of the $^4I_{15/2}$ manifold, where an effective spin-$1/2$ system is valid for identifying the features of the resultant spectra:

\begin{equation}
    H = \mu_B B g S + S A I
    \label{zeeman}
\end{equation}

In the effective spin-$1/2$ Hamiltonian $H$, the first term covers the Zeeman splitting of the electron spin states $S$ proportional to the Bohr magneton $\mu_B$, the applied magnetic field $B$, and the effective $g$-factor produced by local structure. The second term accounts for the hyperfine interaction between the electron spin $S$ and the nuclear spin $I$ for Er isotopes with a non-zero nuclear spin, governed by the hyperfine splitting tensor $A$.

Figure~\ref{fig:fig3}(a) shows a representative example of an EPR spectrum obtained from the 3 ppm Er, 940 nm thick \ch{CeO2} sample.  We identify a primary resonance peak near 100 mT surrounded by a set of lower-intensity resonance peaks. The primary peak arises from the absorption of the \ch{Er^{3+}} electron spin transition for the 77\%{} of naturally abundant nuclear spin $I=0$ Er isotopes (primarily \ch{^{166}Er}, \ch{^{168}Er}, and \ch{^{170}Er}). The lower-intensity peaks arise from the hyperfine interaction between the electron spins and the remaining naturally abundant nuclear spin $I=7/2$ isotope (\ch{^{167}Er}), which yields eight hyperfine peaks. Seven hyperfine peaks are easily identifiable adjacent to the primary peak; the eighth hyperfine peak is obscured by the primary peak \cite{antuzevics2020epr}.

The measured EPR spectrum (Figure~\ref{fig:fig3}(a), blue dots) is fitted (Figure~\ref{fig:fig3}(a), magenta dashed line) to the energy structure defined by Equation~\ref{zeeman} to extract effective $g$-values, hyperfine parameters $A$, and EPR linewidths. Resonance peaks are described with first derivatives of Lorentzians, and the hyperfine peak locations are identified accounting for second-order perturbation in nuclear spin \cite{rieger2007electron}. We extract an effective value $g = 6.812(5)$ and a hyperfine splitting of $A = 687(1)$ MHz for the displayed sample.  This $g$-factor is consistent with theoretical study of \ch{Er^{3+}} residing in a cubic crystal field symmetry \cite{Ammerlaan2001}, and is additionally consistent with experimental study of \ch{Er:CeO2} in bulk and nanocrystal form \cite{Ammerlaan2001, Rakhmatullin2014}.  Based on the cubic symmetry sites available in \ch{CeO2} and the comparable size of the \ch{Ce^{4+}} and \ch{Er^{3+}} ions (0.97 \AA{} and 1.004 \AA{} ionic radii respectively for coordination number 8 \cite{shannon1976revised}), we note that the Er ion likely substitutes into the \ch{Ce} site \cite{minervini1999defect, bao2008structure} under this growth method.

The broadening of resonance peaks in EPR may result from a variety of factors, including magnetic dipole-dipole interactions (e.g. Er-Er) and strain due to defects (e.g. threading dislocations, vacancies, unintentional dopants).  Focusing on the nuclear spin zero peak, we find that the EPR linewidth increases linearly with \ch{Er^{3+}} doping, as shown in Figure~\ref{fig:fig3}(b) for a series of \ch{Er:CeO2}/Si samples (740-940 nm thick, see S.I. for a table of sample details).  A linear increase in linewidth with doping concentration may be associated with broadening due to magnetic dipole-dipole interactions between spins \cite{Geschwind1972EPR}, but we find that broadening due solely to the concentration of \ch{Er^{3+}} ions, $\Gamma_{\text{d-d}}$, would have the values $\Gamma_{\text{d-d}}(\text{2 ppm Er}) = 0.2$ MHz and $\Gamma_{\text{d-d}}(\text{132 ppm Er}) = 15.2$ MHz.  Both of these values are significantly less than the observed linewidths from EPR measurement of those doping levels of 251 MHz and 1438 MHz respectively.  The reason for this discrepancy remains unclear, but suggests that there are other potential dopant-driven broadening mechanisms at play.

\subsection{Effect of erbium concentration on optical characteristics}

\begin{figure}[t]
  \centering
  \includegraphics[width=0.85\linewidth]{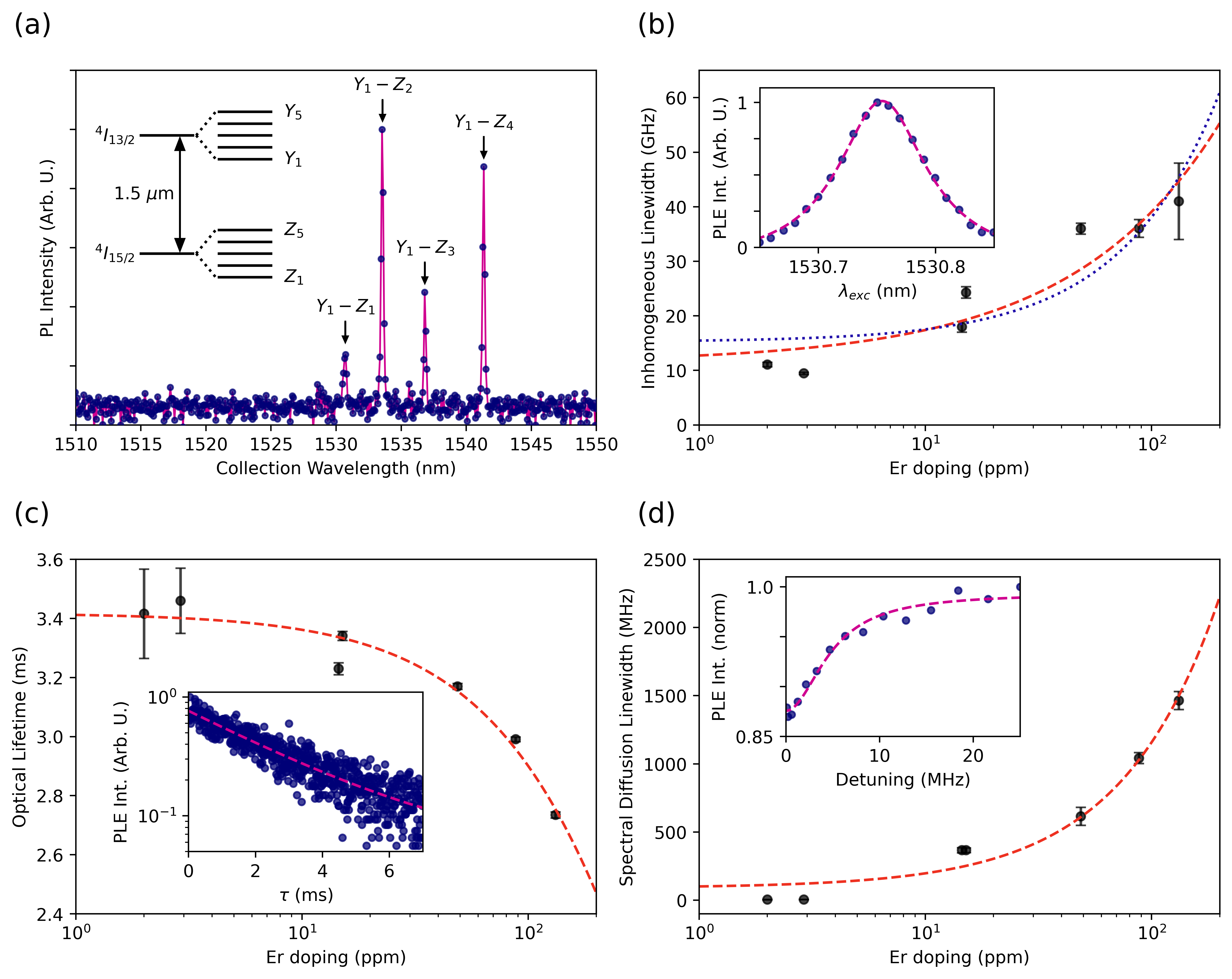}
  \caption{Optical study of \ch{Er:CeO2} thin films on Si, with measurements performed at 3.5 K.  The Er doping dependence fits in (b)-(d) are discussed in detail in the main text.  (a) PL spectrum of a 3 ppm Er, 940 nm thick sample excited by 1473 nm light.  Four transitions are found: $Y_1$ to $Z_{1-4}$.  The magenta line is shown to guide the eye.  (b) Inhomogeneous linewidth of the $Z_1-Y_1$ transition varies with Er concentration (black dots), where each point is the FWHM extracted from the $Z_1-Y_1$ PLE peak (inset, blue dots) via a Lorentzian fit (inset, magenta line).  Power law (red dashed line) and linear (blue dotted line) fits capture the general trend.  (c) Optical excited state lifetime at the $Z_1-Y_1$ transition varies with Er concentration (black dots), where each point is the time constant taken from an optical decay signal (inset, blue dots) by an exponential fit (inset, magenta line).  The dependence of the lifetime on Er doping is fitted to the Inokuti-Hiroyama model (red dashed line).  (d) Spectral diffusion linewidth at the $Z_1-Y_1$ transition varies with Er concentration (black circles), where each point is the HWHM extracted from a TSHB measurement (inset, blue dots) via a Lorentzian fit (inset, magenta curve).  A linear fit (red line) captures doping dependence.}
  \label{fig:fig4}
  \rule{\textwidth}{1pt}
\end{figure}

We study the optical inhomogeneous linewidth $\Gamma_{\text{inh}}$ of the \ch{Er:CeO2} films to characterize the ability to address transitions; the optical spectral diffusion limited linewidth $\Gamma_{\text{SD}}$ as a metric on the optical transition coherence; and the optical excited state lifetime $T_1$ to identify one of the significant time scales of an optical memory interface.

The 11 electrons present in the 4f shell of \ch{Er^{3+}} lead to a ground state electronic configuration of $^4I_{15/2}$ (referred to as Z) and a first excited state electronic configuration of $^4I_{13/2}$ (referred to as Y).  These states split into 5 $Z$ levels and 5 $Y$ levels due to the cubic point symmetry of the host \ch{CeO2} structure \cite{wybourne1965spectroscopic}, which we confirmed by EPR. A diagram of the crystal field-split level structure is shown in Figure~\ref{fig:fig4}(a).

Figure~\ref{fig:fig4}(a) shows the PL spectrum of the 3 ppm Er, 940 nm thick \ch{CeO2} sample excited with 1473 nm light, to the top of the $^4I_{13/2}$ states.  After excitation, a fast non-radiative decay process moves the excited population to $Y_1$ \cite{wolf1976progress}.  Radiative decay from $Y_1$ to the $Z$ levels allows us to observe four $Y_1-Z_i$ transitions, consistent with the maximum five $Z$ levels allowed by cubic symmetry.  The highest energy transition, in this case $Y_1-Z_1$, is found to be at 1530.74(5) nm.  We note that a complete level assignment of all crystal field levels is beyond the scope of this work, but that complete analysis will be published in a forthcoming work.

For additional optical characterization, we focus on the $Z_1-Y_1$ transition due to its technological relevance at low temperature, with the readily accessible spin interface in $Z_1$ as discussed in the EPR section and the absence of the non-radiative processes found in $Y_{>1}$.  We probe the $Z_1-Y_1$ transition with higher spectral resolution using PLE, and examine its inhomogeneous linewidth as a function of Er doping density (Figure~\ref{fig:fig4}(b), black dots), using samples with thicknesses of 740-940 nm (see S.I. for a table of sample details).  The inhomogeneous spread of the absorption line, which ranges from 9.5(2) at 3 ppm Er to 41(7) GHz at 132 ppm Er, may be influenced by the presence of fluctuating electric fields caused by charged defects or strain in the vicinity of the optically active \ch{Er^{3+}} sites.  Such defects can include other \ch{Er^{3+}} ions themselves (since, for example, the aliovalent \ch{Er^{3+}} on a \ch{Ce^{4+}} site will result in a negatively charged point defect \ch{Er_{Ce}^'}, per Kr\"oger-Vink notation \cite{kroger1956relations}), charge compensating defects (e.g. positively charged oxygen vacancies) that are created as a result of the \ch{Er^{3+}} defects to maintain charge neutrality \cite{minervini1999defect}, and ``grown-in'' imperfections during crystal growth.

Taking these things into account, the inhomogeneous linewidth ($\Gamma_{\text{inh}}$) depends on the nature of the interaction between surrounding defects and the optically active emitters \cite{stoneham1969shapes}.  In one scenario, the interaction energy between the emitter (\ch{Er^{3+}}) and a nearby charged defect varies as $\sim1/R^2$, where $R$ is the defect-to-emitter distance, and one expects $\Gamma_{\text{inh}} \propto n^{2/3}$, where $n$ is the defect density \cite{stoneham1969shapes}.  Alternatively, the effects of strain, random electric field gradients, or dipole-dipole interactions -- all with interaction energy $\sim1/R^3$ -- result in a linear dependence $\Gamma_{\text{inh}} \propto n$.  Based on these models, we can describe the generic behavior of the linewidth to be:

\begin{equation}
    \Gamma_{\text{inh}}(n) = a + b n^{2/3} + c n
    \label{inh}
\end{equation}

In Figure~\ref{fig:fig4}(b), we assume for this equation that $n\sim n_{\text{Er}}$, where $n_{\text{Er}}$ is the concentration of \ch{Er_{Ce}'} defects.  Fitting Equation~\ref{inh} to the measured dependence of the inhomogeneous linewidth upon the Er doping concentration (red dashed curve), we find that the $b$ parameter dominates while the linear $c$ parameter goes to zero.  To specifically test the linear dependence case as well, we force a linear fit by setting $b=0$ (blue dotted curve).  We find that both fits capture the generic trend, though $n_{\text{Er}}^{2/3}$ yields a slightly better fit.  Based on this result we conclude that the Er-defect interaction energy may be either $\sim1/R^2$ (charge-dipole) or $\sim1/R^3$ (strain or second order Stark effect) according to Stoneham's analysis \cite{stoneham1969shapes}, but no definite inference of the defect's nature can be made at this stage.  We make note that the $\sim1/R^2$ dependence requires presence of a static dipole formed in the excited Er that is unexpected in the centrosymmetric \ch{CeO2} lattice, and may exist only due to lattice distortions presented by microstructural imperfections.  Finally, we find that there is $\sim$10 GHz residual inhomogeneous broadening even at low Er concentrations of 2-3 ppm.  This is most likely a consequence of grown-in crystalline imperfections during the thin film growth.

Continuing our optical study, we find that the optical excited state lifetime $T_1$ decreases with increasing \ch{Er^{3+}} density as shown in Figure~\ref{fig:fig4}(c) (black dots), from 3.5(1) ms at 2-3 ppm to 2.74(1) ms at 132 ppm.  This may result from the opening up of additional nonradiative pathways with increased density of emitters and defects. A basic understanding may be obtained by applying the Inokuti-Hiroyama theory \cite{inokuti1965influence}, which establishes enhancement of the decay rate as a function of the density of the surrounding defects $n$ that can quench the excitation via energy transfer processes.  Again assuming $n\sim n_{\text{Er}}$, we find the measured behavior can be fit adequately with the Inokuti-Hiroyama theory assuming electric dipole-dipole interactions with the intrinsic lifetime and critical concentration as free parameters, the result of which is shown in Figure~\ref{fig:fig4}(c) (red dashed line).  Additional details on this analysis are presented in the S.I.  We note that these results are consistent with the quenching centers being compensating defects whose concentration is dependent on the \ch{Er^{3+}} doping, although exact nature of these quenching centers can not be inferred at this point.

Finally, Figure~\ref{fig:fig4}(d) shows the dependence of the spectral diffusion-limited homogeneous linewidth, henceforth referred to as the spectral diffusion linewidth $\Gamma_\text{SD}$, as a function of Er doping (black dots).  The varying mean distance of dipole-dipole interactions between excited \ch{Er^{3+}} ions \cite{liu2006spectroscopic} results in a concentration dependence in the spectral diffusion linewidth.  We confirm that this is reflected in the spectral diffusion linewidth increasing with doping, from 4.8(3) MHz at 2 ppm to 1465(66) MHz at 132 ppm.  For a large ensemble of \ch{Er^{3+}} ions, each optically active ion experiences random instantaneous spectral diffusion (ISD) \cite{graf1998photon} caused by the dipole-dipole interaction with nearby excited Er ions.  This results in additional dephasing manifest in the spectral diffusion, which for ISD-caused broadening should be linear in the density of the excited ions \cite{graf1998photon} indicating $\Gamma_\text{SD} = a + b n_{\text{Er}}$ (red dashed line in Figure~\ref{fig:fig4}(d)).  We emphasize the significant correlation between the doping density and spectral diffusion linewidth, particularly since at the lowest doping levels of 2-3 ppm we see spectral diffusion linewidths of $\sim$5 MHz for a millisecond-timescale TSHB measurement conducted at 3.5 K.

\subsection{Effect of annealing on \ch{Er:CeO2} optical characteristics}

\begin{figure}[t]
  \centering
  \includegraphics[width=0.75\linewidth]{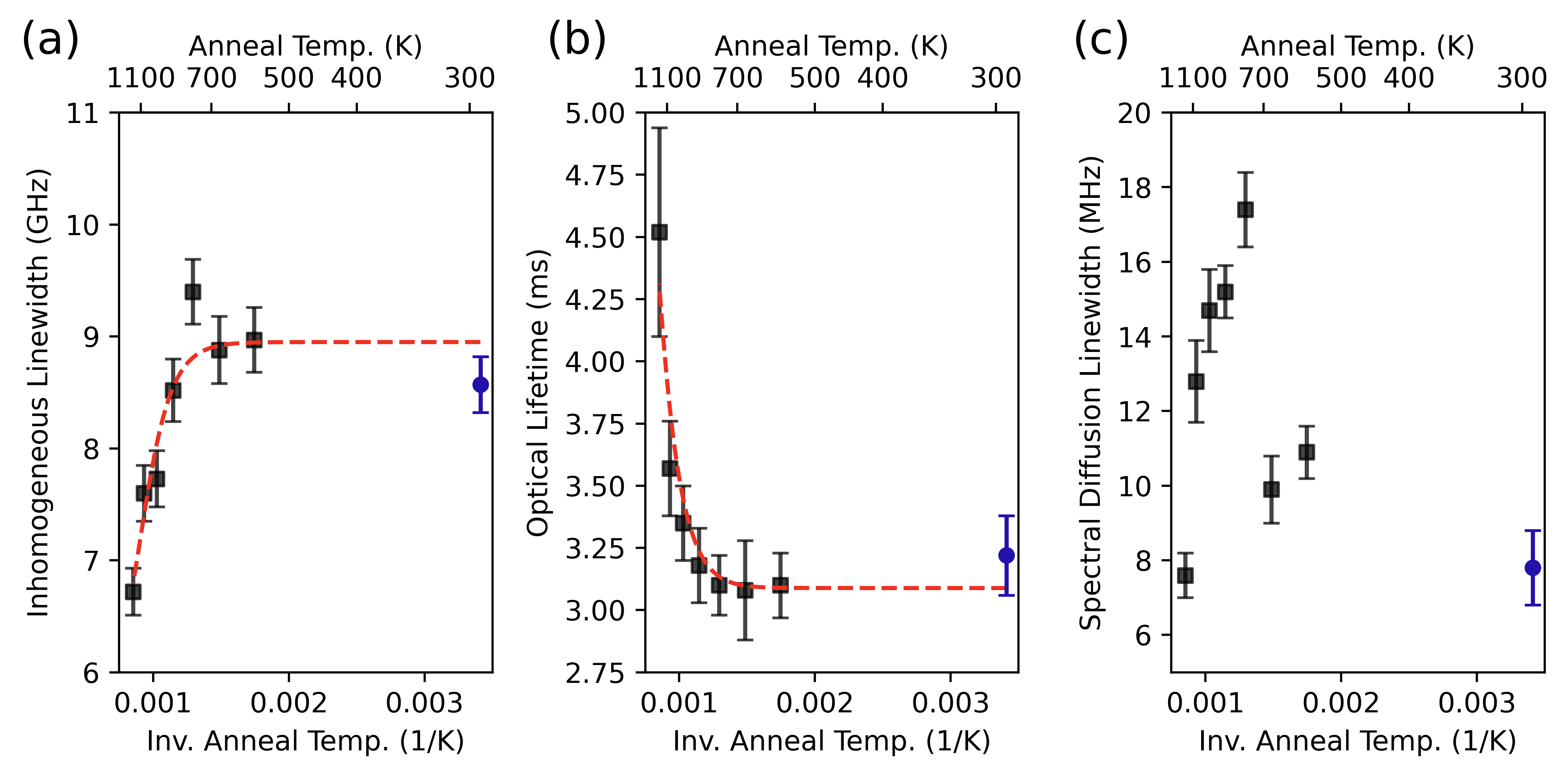}
  \caption{Annealing study of a 3 ppm Er, 200 nm thick \ch{CeO2} thin film on Si via 12-hour anneals in \ch{O2}/\ch{Ar} at 1 atmosphere, measuring optical properties as a function of annealing temperature.  Optical measurements are performed at 3.5 K.  Black squares indicate annealed samples, blue circles indicate the un-annealed as-grown sample as the built-in reference.  Kinetic model extensions to the inhomogeneous linewidth and optical lifetime models (red dashed curves) are described in the text. (a) Inhomogeneous linewidths  before and after annealing.  (b) Optical lifetime of the $Z_1-Y_1$ transition before and after annealing.  (c) Spectral diffusion linewidth of the $Z_1-Y_1$ transition before and after annealing.}
  \label{fig:fig5}
  \rule{\textwidth}{1pt}
\end{figure}

Though we are careful not to assume the nature of the defects leading to broadening and quenching in the doping series results, we speculate that these defects may be partially mitigated by post-growth annealing.  Oxygen vacancies, for example, are a common side-effect of MBE processes due to the high-temperature, low-pressure environment used for growth and may be removed by annealing.  Annealing is also a common step when processing samples produced by other means, e.g. after ion implantation of Er into bulk \cite{stevenson2022erbium}, and so is a useful point of comparison.

To study the effect of annealing we studied a 200 nm thick, 3 ppm \ch{Er:CeO2} film on Si for 12 hours in 1 atmosphere of 20\%{} \ch{O2}/\ch{Ar}, at different temperatures up to 900 \celsius{} (film roughening occurred beyond this temperature).  Figures~\ref{fig:fig5}(a-c) show the dependencies of the PLE-measured inhomogeneous linewidth, radiative lifetime, and spectral diffusion linewidth of the $Z_1-Y_1$ transition as a function of the annealing temperature.  Annealing leads to modest improvements in $\Gamma_{\text{inh}}$ and $T_1$ of 20\%{} and 40\%{} respectively from their as-grown values.  We ascribe these improvements due to the annealing out of ``grown-in" crystal defects in the thin films.  However, the spectral diffusion linewidth worsens at moderate temperatures and returns to the as-grown linewidth at the maximum temperature studied, and the process driving this behavior is unclear.

The trends in inhomogeneous linewidth $\Gamma_{\text{inh}}$ and the excited state lifetime ($T_1$) as a function of annealing temperature ($T$) can be captured by assuming (i) a first-order reaction rate-limited process of thermally activated annihilation of the grown-in defects that affect the optical properties, and (ii) the suitability of the previously described power law relation and the Inokuti-Hirayama approach respectively for the defect concentration dependence upon $\Gamma_{\text{inh}}$ and $T_1$.  Details of these models are given in the S.I.  We find that an Arrhenius-like activation energy $E_A$ in the range of $0.65-0.75$ eV for the temperature-dependent first-order reaction rate constant leads to good fits for both $\Gamma_{\text{inh}}$ (Figure~\ref{fig:fig5}(a), red dashed line) and $T_1$ (Figure~\ref{fig:fig5}(b), red dashed line).  This points to a density of grown-in, optically relevant defects that are being annihilated via thermally activated processes.

\section{Conclusion}

The \ch{Er:CeO2}/Si system presents an attractive combination of benefits, as it is an ideal host oxide for a spin defect with its very low nuclear noise environment, and has low lattice mismatch for epitaxial growth on silicon.  In this work, we have carried out a detailed microstructural and optical study of MBE-grown epitaxial \ch{Er:CeO2}/Si in the 2-132 ppm Er doping range, yielding results relevant for development in quantum coherent device applications. We establish a baseline for this material in the context of key metrics for rare-earth doped oxide systems: as-grown films at 2-3 ppm Er doping show EPR linewidths as narrow as $245(1)$ MHz, optical inhomogeneous linewidths down to $9.5(2)$ GHz, an optical excited state lifetime as long as $3.5(1)$ ms, and a spectral diffusion-limited homogeneous linewidth as narrow as $4.8(3)$ MHz.  Annealing to 900 \celsius{} improves the optical inhomogeneous linewidth and excited state lifetime by a modest 20\%{} and 40 \%{} respectively, yielding an inhomogeneous linewidth as narrow as $6.7(2)$ GHz and an optical excited state lifetime as long as $4.5(4)$ ms.

In studying the doping dependence of the optical parameters as a function of Er doping, we show that the functional dependence is consistent with a charge dipole-based interaction model between the Er emitters.  Overall the optical linewidths for the thin films are broader than corresponding linewidths in high-quality bulk samples doped with Er, likely due to the larger number of grown-in defects such as threading dislocations formed during thin film growth.  As such, our future research will target reducing defect densities via growth process optimization.  We also note that the narrow $\sim5$ MHz spectral diffusion linewidths at 2-3 ppm Er doping are sufficiently low to begin exploring measurement techniques such as photon echo, which will allow us to directly probe optical coherence of \ch{Er^{3+}} in \ch{CeO2}.

\section*{Acknowledgments}

The authors would like to thank Dr. Jasleen K. Bindra for assistance with EPR measurements.  This work is primarily supported by Q-NEXT, a U.S. Department of Energy Office of Science National Quantum Information Science Research Centers under Award Number DE-FOA-0002253.  Use of the Center for Nanoscale Materials, an Office of Science User Facility, use of the Advanced Photon Source at Argonne National Laboratory, and the EPR work in the Chemical Sciences and Engineering Division were supported by the U.S. Department of Energy, Office of Science, Office of Basic Energy Sciences, under Contract No. DE-AC02-06CH11357. This work also made use of the Jerome B. Cohen X-Ray Diffraction Facility supported by the MRSEC program of the National Science Foundation (DMR-2308691) at the Materials Research Center of Northwestern University and the Soft and Hybrid Nanotechnology Experimental (SHyNE) Resource (NSF ECCS-1542205).

\clearpage

\bibliography{main}  

\begin{thebibliography}{55}
\providecommand{\natexlab}[1]{#1}
\providecommand{\url}[1]{\texttt{#1}}
\expandafter\ifx\csname urlstyle\endcsname\relax
  \providecommand{\doi}[1]{doi: #1}\else
  \providecommand{\doi}{doi: \begingroup \urlstyle{rm}\Url}\fi

\bibitem[Awschalom et~al.(2022)Awschalom, Bernien, Brown, Clerk, Chitambar, Dibos, Dionne, Eriksson, Fefferman, Fuchs, et~al.]{awschalom2022roadmap}
David~D Awschalom, Hannes Bernien, Rex Brown, Aashish Clerk, Eric Chitambar, Alan Dibos, Jennifer Dionne, Mark Eriksson, Bill Fefferman, Greg~David Fuchs, et~al.
\newblock A roadmap for quantum interconnects.
\newblock Technical report, Argonne National Lab. (ANL), Argonne, IL (United States), 2022.

\bibitem[Awschalom et~al.(2021)Awschalom, Berggren, Bernien, Bhave, Carr, Davids, Economou, Englund, Faraon, Fejer, et~al.]{awschalom2021development}
David Awschalom, Karl~K Berggren, Hannes Bernien, Sunil Bhave, Lincoln~D Carr, Paul Davids, Sophia~E Economou, Dirk Englund, Andrei Faraon, Martin Fejer, et~al.
\newblock Development of quantum interconnects (quics) for next-generation information technologies.
\newblock \emph{PRX Quantum}, 2\penalty0 (1):\penalty0 017002, 2021.
\newblock \doi{10.1103/PRXQuantum.2.017002}.

\bibitem[Kinos et~al.(2021)Kinos, Hunger, Kolesov, M{\o}lmer, de~Riedmatten, Goldner, Tallaire, Morvan, Berger, Welinski, Karrai, Rippe, Kr{\"{o}}ll, and Walther]{Kinos2021}
Adam Kinos, David Hunger, Roman Kolesov, Klaus M{\o}lmer, Hugues de~Riedmatten, Philippe Goldner, Alexandre Tallaire, Loic Morvan, Perrine Berger, Sacha Welinski, Khaled Karrai, Lars Rippe, Stefan Kr{\"{o}}ll, and Andreas Walther.
\newblock {Roadmap for Rare-earth Quantum Computing}.
\newblock \emph{arXiv preprint 2103.15743}, pages 1--47, 2021.

\bibitem[Liu and Jacquier(2006)]{liu2006spectroscopic}
Guokui Liu and Bernard Jacquier.
\newblock \emph{Spectroscopic properties of rare earths in optical materials}, volume~83.
\newblock Springer Science \& Business Media, 2006.
\newblock ISBN 9783642062834.

\bibitem[Wolfowicz et~al.(2021)Wolfowicz, Heremans, Anderson, Kanai, Seo, Gali, Galli, and Awschalom]{Wolfowicz2021}
Gary Wolfowicz, F.~Joseph Heremans, Christopher~P. Anderson, Shun Kanai, Hosung Seo, Adam Gali, Giulia Galli, and David~D. Awschalom.
\newblock {Quantum guidelines for solid-state spin defects}.
\newblock \emph{Nature Reviews Materials}, 6\penalty0 (10):\penalty0 906--925, 2021.
\newblock \doi{10.1038/s41578-021-00306-y}.

\bibitem[Singh et~al.(2022)Singh, Wolfowicz, Wen, Sullivan, Prakash, Dibos, Awschalom, Heremans, and Guha]{singh2022development}
Manish~Kumar Singh, Gary Wolfowicz, Jianguo Wen, Sean~E Sullivan, Abhinav Prakash, Alan~M Dibos, David~D Awschalom, F~Joseph Heremans, and Supratik Guha.
\newblock Development of a scalable quantum memory platform--materials science of erbium-doped \ch{TiO2} thin films on silicon.
\newblock \emph{arXiv preprint 2202.05376v2}, 2022.

\bibitem[Dibos et~al.(2022)Dibos, Solomon, Sullivan, Singh, Sautter, Horn, Grant, Lin, Wen, Heremans, et~al.]{dibos2022purcell}
Alan~M Dibos, Michael~T Solomon, Sean~E Sullivan, Manish~K Singh, Kathryn~E Sautter, Connor~P Horn, Gregory~D Grant, Yulin Lin, Jianguo Wen, F~Joseph Heremans, et~al.
\newblock Purcell enhancement of erbium ions in \ch{TiO2{}} on silicon nanocavities.
\newblock \emph{Nano Letters}, 22\penalty0 (16):\penalty0 6530--6536, 2022.
\newblock \doi{10.1021/acs.nanolett.2c01561}.

\bibitem[Ji et~al.(2023)Ji, Solomon, Grant, Tanaka, Hua, Wen, Seth, Horn, Masiulionis, Singh, Sullivan, Heremans, Awschalom, Guha, and Dibos]{ji2023nanocavitymediated}
Cheng Ji, Michael~T. Solomon, Gregory~D. Grant, Koichi Tanaka, Muchuan Hua, Jianguo Wen, Sagar~K. Seth, Connor~P. Horn, Ignas Masiulionis, Manish~K. Singh, Sean~E. Sullivan, F.~Joseph Heremans, David~D. Awschalom, Supratik Guha, and Alan~M. Dibos.
\newblock Nanocavity-mediated purcell enhancement of \ch{Er} in \ch{TiO2} thin films grown via atomic layer deposition.
\newblock \emph{arXiv prepring arXiv:2309.13490}, 2023.

\bibitem[Singh et~al.(2020)Singh, Prakash, Wolfowicz, Wen, Huang, Rajh, Awschalom, Zhong, and Guha]{singh2020epitaxial}
Manish~Kumar Singh, Abhinav Prakash, Gary Wolfowicz, Jianguo Wen, Yizhong Huang, Tijana Rajh, David~D Awschalom, Tian Zhong, and Supratik Guha.
\newblock Epitaxial \ch{Er}-doped \ch{Y2O3} on silicon for quantum coherent devices.
\newblock \emph{APL Materials}, 8\penalty0 (3):\penalty0 031111, 2020.
\newblock \doi{10.1063/1.5142611}.

\bibitem[Gupta et~al.(2023)Gupta, Liu, Wang, Huang, and Zhong]{gupta2023towards}
Shobhit Gupta, Shihan Liu, Chao-Fan Wang, Yizhong Huang, and Tian Zhong.
\newblock Towards khz optical linewidth, millisecond spin coherence erbium telecom qubits in epitaxial thin films.
\newblock In \emph{CLEO: Fundamental Science}, pages FTh1A--2. Optica Publishing Group, 2023.
\newblock \doi{10.1364/CLEO_FS.2023.FTh1A.2}.

\bibitem[Le~Dantec et~al.(2021)Le~Dantec, Ran{\v{c}}i{\'c}, Lin, Billaud, Ranjan, Flanigan, Bertaina, Chaneli{\`e}re, Goldner, Erb, et~al.]{le2021twenty}
Marianne Le~Dantec, Milo{\v{s}} Ran{\v{c}}i{\'c}, Sen Lin, Eric Billaud, Vishal Ranjan, Daniel Flanigan, Sylvain Bertaina, Thierry Chaneli{\`e}re, Philippe Goldner, Andreas Erb, et~al.
\newblock Twenty-three--millisecond electron spin coherence of erbium ions in a natural-abundance crystal.
\newblock \emph{Science advances}, 7\penalty0 (51):\penalty0 eabj9786, 2021.
\newblock \doi{sciadv.abj9786}.

\bibitem[Ourari et~al.(2023)Ourari, Dusanowski, Horvath, Uysal, Phenicie, Stevenson, Raha, Chen, Cava, de~Leon, and Thompson]{Ourari2023}
Salim Ourari, {\L}ukasz Dusanowski, Sebastian~P. Horvath, Mehmet~T. Uysal, Christopher~M. Phenicie, Paul Stevenson, Mouktik Raha, Songtao Chen, Robert~J. Cava, Nathalie~P. de~Leon, and Jeff~D. Thompson.
\newblock Indistinguishable telecom band photons from a single er ion in the solid state.
\newblock \emph{Nature}, 620\penalty0 (7976):\penalty0 977--981, Aug 2023.
\newblock ISSN 1476-4687.
\newblock \doi{10.1038/s41586-023-06281-4}.

\bibitem[Ran{\v{c}}i{\'c} et~al.(2018)Ran{\v{c}}i{\'c}, Hedges, Ahlefeldt, and Sellars]{ranvcic2018coherence}
Milo{\v{s}} Ran{\v{c}}i{\'c}, Morgan~P Hedges, Rose~L Ahlefeldt, and Matthew~J Sellars.
\newblock Coherence time of over a second in a telecom-compatible quantum memory storage material.
\newblock \emph{Nature Physics}, 14\penalty0 (1):\penalty0 50--54, 2018.
\newblock \doi{10.1038/nphys4254}.

\bibitem[Horvath et~al.(2023)Horvath, Phenicie, Ourari, Uysal, Chen, Dusanowski, Raha, Stevenson, Turflinger, Cava, et~al.]{horvath2023strong}
Sebastian~P Horvath, Christopher~M Phenicie, Salim Ourari, Mehmet~T Uysal, Songtao Chen, {\L}ukasz Dusanowski, Mouktik Raha, Paul Stevenson, Adam~T Turflinger, Robert~J Cava, et~al.
\newblock Strong purcell enhancement of an optical magnetic dipole transition.
\newblock \emph{arXiv preprint arXiv:2307.03022}, 2023.

\bibitem[Rochman et~al.(2023)Rochman, Xie, Bartholomew, Schwab, and Faraon]{rochman2023microwave}
Jake Rochman, Tian Xie, John~G Bartholomew, KC~Schwab, and Andrei Faraon.
\newblock Microwave-to-optical transduction with erbium ions coupled to planar photonic and superconducting resonators.
\newblock \emph{Nature Communications}, 14\penalty0 (1):\penalty0 1153, 2023.
\newblock \doi{10.1038/s41467-023-36799-0}.

\bibitem[Yu et~al.(2023)Yu, Oser, Da~Prato, Urbinati, {\'A}vila, Zhang, Remy, Marzban, Gr{\"o}blacher, and Tittel]{yu2023frequency}
Yong Yu, Dorian Oser, Gaia Da~Prato, Emanuele Urbinati, Javier~Carrasco {\'A}vila, Yu~Zhang, Patrick Remy, Sara Marzban, Simon Gr{\"o}blacher, and Wolfgang Tittel.
\newblock Frequency tunable, cavity-enhanced single erbium quantum emitter in the telecom band.
\newblock \emph{arXiv preprint arXiv:2304.14685}, 2023.

\bibitem[Weiss et~al.(2021)Weiss, Gritsch, Merkel, and Reiserer]{Weiss21}
Lorenz Weiss, Andreas Gritsch, Benjamin Merkel, and Andreas Reiserer.
\newblock Erbium dopants in nanophotonic silicon waveguides.
\newblock \emph{Optica}, 8\penalty0 (1):\penalty0 40--41, Jan 2021.
\newblock \doi{10.1364/OPTICA.413330}.

\bibitem[Rinner et~al.(2023)Rinner, Burger, Gritsch, Schmitt, and Reiserer]{rinner2023erbium}
Stephan Rinner, Florian Burger, Andreas Gritsch, Jonas Schmitt, and Andreas Reiserer.
\newblock Erbium emitters in commercially fabricated nanophotonic silicon waveguides.
\newblock \emph{Nanophotonics}, 12\penalty0 (17):\penalty0 3455--3462, 2023.
\newblock \doi{10.1515/nanoph-2023-0287}.

\bibitem[Kanai et~al.(2022)Kanai, Heremans, Seo, Wolfowicz, Anderson, Sullivan, Onizhuk, Galli, Awschalom, and Ohno]{kanai2022generalized}
Shun Kanai, F~Joseph Heremans, Hosung Seo, Gary Wolfowicz, Christopher~P Anderson, Sean~E Sullivan, Mykyta Onizhuk, Giulia Galli, David~D Awschalom, and Hideo Ohno.
\newblock Generalized scaling of spin qubit coherence in over 12,000 host materials.
\newblock \emph{Proceedings of the National Academy of Sciences}, 119\penalty0 (15):\penalty0 e2121808119, 2022.
\newblock \doi{10.1073/pnas.2121808119}.

\bibitem[Stoll and Schweiger(2006)]{stoll2006easyspin}
Stefan Stoll and Arthur Schweiger.
\newblock Easyspin, a comprehensive software package for spectral simulation and analysis in epr.
\newblock \emph{Journal of magnetic resonance}, 178\penalty0 (1):\penalty0 42--55, 2006.
\newblock \doi{10.1016/j.jmr.2005.08.013}.

\bibitem[Yoshimoto et~al.(1990)Yoshimoto, Nagata, Tsukahara, and Koinuma]{yoshimoto1990situ}
Mamoru Yoshimoto, Hirotoshi Nagata, Tadashi Tsukahara, and Hideomi Koinuma.
\newblock In situ rheed observation of \ch{CeO2} film growth on \ch{Si} by laser ablation deposition in ultrahigh-vacuum.
\newblock \emph{Japanese journal of applied physics}, 29\penalty0 (7A):\penalty0 L1199, 1990.
\newblock \doi{10.1143/JJAP.29.L1199}.

\bibitem[Nagata et~al.(1992)Nagata, Yoshimoto, Koinuma, Min, and Haga]{nagata1992type}
Hirotoshi Nagata, Mamoru Yoshimoto, Hideomi Koinuma, Eungi Min, and Nobuhiko Haga.
\newblock Type-b epitaxial growth of \ch{CeO2} thin film on \ch{Si}(111) substrate.
\newblock \emph{Journal of crystal growth}, 123\penalty0 (1-2):\penalty0 1--4, 1992.
\newblock \doi{10.1016/0022-0248(92)90004-3}.

\bibitem[Nishikawa et~al.(2002)Nishikawa, Fukushima, Yasuda, Nakayama, and Ikegawa]{nishikawa2002electrical}
Yukie Nishikawa, Noburu Fukushima, Naoki Yasuda, Kohei Nakayama, and Sumio Ikegawa.
\newblock Electrical properties of single crystalline \ch{CeO2} high-k gate dielectrics directly grown on \ch{Si}(111).
\newblock \emph{Japanese journal of applied physics}, 41\penalty0 (4S):\penalty0 2480, 2002.
\newblock \doi{10.1143/JJAP.41.2480}.

\bibitem[Barth et~al.(2016)Barth, Laffon, Olbrich, Ranguis, Parent, and Reichling]{barth2016perfectly}
C~Barth, C~Laffon, R~Olbrich, A~Ranguis, Ph~Parent, and M~Reichling.
\newblock A perfectly stoichiometric and flat \ch{CeO2} (111) surface on a bulk-like ceria film.
\newblock \emph{Scientific Reports}, 6\penalty0 (1):\penalty0 21165, 2016.
\newblock \doi{10.1038/srep21165}.

\bibitem[Inaba et~al.(2018)Inaba, Tawara, Omi, Yamamoto, and Gotoh]{inaba2018epitaxial}
Tomohiro Inaba, Takehiko Tawara, Hiroo Omi, Hideki Yamamoto, and Hideki Gotoh.
\newblock Epitaxial growth and optical properties of \ch{Er}-doped \ch{CeO2} on \ch{Si}(111).
\newblock \emph{Optical Materials Express}, 8\penalty0 (9):\penalty0 2843--2849, 2018.
\newblock \doi{10.1364/OME.8.002843}.

\bibitem[Tawara et~al.(2017)Tawara, Mariani, Shimizu, Omi, Adachi, and Gotoh]{tawara2017effect}
Takehiko Tawara, Giacomo Mariani, Kaoru Shimizu, Hiroo Omi, Satoru Adachi, and Hideki Gotoh.
\newblock Effect of isotopic purification on spectral-hole narrowing in \ch{^{167}Er^{3+}} hyperfine transitions.
\newblock \emph{Applied Physics Express}, 10\penalty0 (4):\penalty0 042801, 2017.
\newblock \doi{10.7567/APEX.10.042801}.

\bibitem[Thiel et~al.(2011)Thiel, B{\"o}ttger, and Cone]{thiel2011rare}
Charles~W Thiel, Thomas B{\"o}ttger, and RL~Cone.
\newblock Rare-earth-doped materials for applications in quantum information storage and signal processing.
\newblock \emph{Journal of luminescence}, 131\penalty0 (3):\penalty0 353--361, 2011.
\newblock \doi{10.1016/j.jlumin.2010.12.015}.

\bibitem[Wang et~al.(2022)Wang, Yang, Shen, Fu, Xu, Cone, Thiel, and Tang]{wang2022er}
Sihao Wang, Likai Yang, Mohan Shen, Wei Fu, Yuntao Xu, Rufus~L Cone, Charles~W Thiel, and Hong~X Tang.
\newblock \ch{Er:LiNbO3} with high optical coherence enabling optical thickness control.
\newblock \emph{Physical Review Applied}, 18\penalty0 (1):\penalty0 014069, 2022.
\newblock \doi{10.1103/PhysRevApplied.18.014069}.

\bibitem[Mayer et~al.(2007)Mayer, Giannuzzi, Kamino, and Michael]{mayer2007tem}
Joachim Mayer, Lucille~A Giannuzzi, Takeo Kamino, and Joseph Michael.
\newblock Tem sample preparation and fib-induced damage.
\newblock \emph{MRS bulletin}, 32\penalty0 (5):\penalty0 400--407, 2007.
\newblock \doi{10.1557/mrs2007.63}.

\bibitem[Narayanan et~al.(2002)Narayanan, Guha, Copel, Bojarczuk, Flaitz, and Gribelyuk]{narayanan2002interfacial}
V~Narayanan, S~Guha, M~Copel, NA~Bojarczuk, PL~Flaitz, and M~Gribelyuk.
\newblock Interfacial oxide formation and oxygen diffusion in rare earth oxide--silicon epitaxial heterostructures.
\newblock \emph{Applied physics letters}, 81\penalty0 (22):\penalty0 4183--4185, 2002.
\newblock \doi{10.1063/1.1524692}.

\bibitem[Thole et~al.(1985)Thole, Van~der Laan, Fuggle, Sawatzky, Karnatak, and Esteva]{thole19853d}
BT~Thole, G~Van~der Laan, JC~Fuggle, GA~Sawatzky, RC~Karnatak, and J-M Esteva.
\newblock 3d x-ray-absorption lines and the 3d$^9$ 4f$^{n+1}$ multiplets of the lanthanides.
\newblock \emph{Physical Review B}, 32\penalty0 (8):\penalty0 5107, 1985.
\newblock \doi{10.1103/PhysRevB.32.5107}.

\bibitem[Paidi et~al.(2019)Paidi, Brewe, Freeland, Roberts, and van Lierop]{paidi2019role}
Vinod~K Paidi, Dale~L Brewe, John~W Freeland, Charles~A Roberts, and Johan van Lierop.
\newblock Role of \ch{Ce} 4f hybridization in the origin of magnetism in nanoceria.
\newblock \emph{Physical Review B}, 99\penalty0 (18):\penalty0 180403, 2019.
\newblock \doi{10.1103/PhysRevB.99.180403}.

\bibitem[Chen et~al.(2012)Chen, Tsai, Huang, Yan, Huang, Gloter, Chen, Lin, Chen, and Dong]{chen2012concentration}
Shih-Yun Chen, Chi-Hang Tsai, Mei-Zi Huang, Der-Chung Yan, Tzu-Wen Huang, Alexandre Gloter, Chi-Liang Chen, Hong-Ji Lin, Chien-Te Chen, and Chung-Li Dong.
\newblock Concentration dependence of oxygen vacancy on the magnetism of \ch{CeO2} nanoparticles.
\newblock \emph{The Journal of Physical Chemistry C}, 116\penalty0 (15):\penalty0 8707--8713, 2012.
\newblock \doi{10.1021/jp2065634}.

\bibitem[Garvie and Buseck(1999)]{garvie1999ce4ce3}
L.A.J. Garvie and P.R. Buseck.
\newblock Determination of ce4+/ce3+ in electron-beam-damaged ceo2 by electron energy-loss spectroscopy.
\newblock \emph{Journal of Physics and Chemistry of Solids}, 60\penalty0 (12):\penalty0 1943--1947, 1999.
\newblock ISSN 0022-3697.
\newblock \doi{10.1016/S0022-3697(99)00218-8}.

\bibitem[Hayakawa et~al.(2016)Hayakawa, Egashira, Arakawa, Ito, Sarugaku, Ando, and Terasaki]{hayakawa2016x}
Tetsuichiro Hayakawa, Kazuhiro Egashira, Masashi Arakawa, Tomonori Ito, Shun Sarugaku, Kota Ando, and Akira Terasaki.
\newblock X-ray absorption spectroscopy of \ch{Ce2O3^+} and \ch{Ce2O5^+} near \ch{Ce} m-edge.
\newblock \emph{Journal of Physics B: Atomic, Molecular and Optical Physics}, 49\penalty0 (7):\penalty0 075101, 2016.
\newblock \doi{10.1088/0953-4075/49/7/075101}.

\bibitem[Antuzevics(2020)]{antuzevics2020epr}
Andris Antuzevics.
\newblock Epr characterization of erbium in glasses and glass ceramics.
\newblock \emph{Low Temperature Physics}, 46\penalty0 (12):\penalty0 1149--1153, 2020.
\newblock \doi{10.1063/10.0002465}.

\bibitem[Rieger(2007)]{rieger2007electron}
Philip Rieger.
\newblock \emph{Electron spin resonance: analysis and interpretation}.
\newblock Royal Society of Chemistry, 2007.
\newblock ISBN 9780854043552.

\bibitem[Ammerlaan and {De Maat-Gersdorf}(2001)]{Ammerlaan2001}
Cornelis~A.J. Ammerlaan and I.~{De Maat-Gersdorf}.
\newblock {Zeeman splitting factor of the \ch{Er^{3+}} ion in a crystal field}.
\newblock \emph{Applied Magnetic Resonance}, 21\penalty0 (1):\penalty0 13--33, 2001.
\newblock ISSN 09379347.
\newblock \doi{10.1007/BF03162436}.

\bibitem[Rakhmatullin et~al.(2014)Rakhmatullin, Kurkin, Pavlov, and Semashko]{Rakhmatullin2014}
R.~M. Rakhmatullin, I.~N. Kurkin, V.~V. Pavlov, and V.~V. Semashko.
\newblock {EPR, optical, and dielectric spectroscopy of Er-doped cerium dioxide nanoparticles}.
\newblock \emph{Physica Status Solidi (B) Basic Research}, 251\penalty0 (8):\penalty0 1545--1551, 2014.
\newblock ISSN 15213951.
\newblock \doi{10.1002/pssb.201451116}.

\bibitem[Shannon(1976)]{shannon1976revised}
Robert~D Shannon.
\newblock Revised effective ionic radii and systematic studies of interatomic distances in halides and chalcogenides.
\newblock \emph{Acta crystallographica section A: crystal physics, diffraction, theoretical and general crystallography}, 32\penalty0 (5):\penalty0 751--767, 1976.
\newblock \doi{10.1107/S0567739476001551}.

\bibitem[Minervini et~al.(1999)Minervini, Zacate, and Grimes]{minervini1999defect}
Licia Minervini, Matthew~O Zacate, and Robin~W Grimes.
\newblock Defect cluster formation in \ch{M2O3}-doped \ch{CeO2}.
\newblock \emph{Solid State Ionics}, 116\penalty0 (3-4):\penalty0 339--349, 1999.
\newblock \doi{10.1016/S0167-2738(98)00359-2}.

\bibitem[Bao et~al.(2008)Bao, Chen, Fang, Jiang, and Huang]{bao2008structure}
Huizhi Bao, Xin Chen, Jun Fang, Zhiquan Jiang, and Weixin Huang.
\newblock Structure-activity relation of \ch{Fe2O3}-\ch{CeO2} composite catalysts in \ch{CO} oxidation.
\newblock \emph{Catalysis letters}, 125:\penalty0 160--167, 2008.
\newblock \doi{10.1007/s10562-008-9540-3}.

\bibitem[Geschwind(1972)]{Geschwind1972EPR}
S~Geschwind, editor.
\newblock \emph{Electron Paramagnetic Resonance}.
\newblock Plenum Press, New York, 1972.
\newblock ISBN 0306305801.

\bibitem[Wybourne and Meggers(1965)]{wybourne1965spectroscopic}
Brian~G Wybourne and William~F Meggers.
\newblock \emph{Spectroscopic properties of rare earths}.
\newblock American Institute of Physics, 1965.
\newblock ISBN 9780470965078.

\bibitem[Wolf and Dainty(1976)]{wolf1976progress}
E.~Wolf and J.C. Dainty.
\newblock \emph{Progress in Optics}.
\newblock Number v. 14 in Progress in optics. North-Holland Publishing Company, 1976.
\newblock ISBN 9780720415148.

\bibitem[Kr{\"o}ger and Vink(1956)]{kroger1956relations}
FA~Kr{\"o}ger and HJ~Vink.
\newblock Relations between the concentrations of imperfections in crystalline solids.
\newblock In \emph{Solid state physics}, volume~3, pages 307--435. Elsevier, 1956.
\newblock \doi{10.1016/S0081-1947(08)60135-6}.

\bibitem[Stoneham(1969)]{stoneham1969shapes}
AM~Stoneham.
\newblock Shapes of inhomogeneously broadened resonance lines in solids.
\newblock \emph{Reviews of Modern Physics}, 41\penalty0 (1):\penalty0 82, 1969.
\newblock \doi{10.1103/RevModPhys.41.82}.

\bibitem[Inokuti and Hirayama(1965)]{inokuti1965influence}
Mitio Inokuti and Fumio Hirayama.
\newblock Influence of energy transfer by the exchange mechanism on donor luminescence.
\newblock \emph{The journal of chemical physics}, 43\penalty0 (6):\penalty0 1978--1989, 1965.
\newblock \doi{10.1063/1.1697063}.

\bibitem[Graf et~al.(1998)Graf, Renn, Zumofen, and Wild]{graf1998photon}
Felix~R Graf, Alois Renn, Gert Zumofen, and Urs~P Wild.
\newblock Photon-echo attenuation by dynamical processes in rare-earth-ion-doped crystals.
\newblock \emph{Physical Review B}, 58\penalty0 (9):\penalty0 5462, 1998.
\newblock \doi{10.1103/PhysRevB.58.5462}.

\bibitem[Stevenson et~al.(2022)Stevenson, Phenicie, Gray, Horvath, Welinski, Ferrenti, Ferrier, Goldner, Das, Ramesh, et~al.]{stevenson2022erbium}
Paul Stevenson, Christopher~M Phenicie, Isaiah Gray, Sebastian~P Horvath, Sacha Welinski, Austin~M Ferrenti, Alban Ferrier, Philippe Goldner, Sujit Das, Ramamoorthy Ramesh, et~al.
\newblock Erbium-implanted materials for quantum communication applications.
\newblock \emph{Physical Review B}, 105\penalty0 (22):\penalty0 224106, 2022.
\newblock \doi{10.1103/PhysRevB.105.224106}.

\bibitem[Kern(2018)]{kern2018cleaning}
Werner Kern.
\newblock Chapter 1 - overview and evolution of silicon wafer cleaning technology.
\newblock In Karen~A. Reinhardt and Werner Kern, editors, \emph{Handbook of Silicon Wafer Cleaning Technology (Third Edition)}, pages 3--85. William Andrew Publishing, third edition edition, 2018.
\newblock ISBN 9780323510844.
\newblock \doi{10.1016/B978-0-323-51084-4.00001-0}.
\newblock URL \url{https://www.sciencedirect.com/science/article/pii/B9780323510844000010}.

\bibitem[Bertness(2000)]{bertness2000smart}
Kristine~A Bertness.
\newblock Smart pyrometry for combined sample temperature and reflectance measurements in molecular-beam epitaxy.
\newblock \emph{Journal of Vacuum Science \& Technology B: Microelectronics and Nanometer Structures Processing, Measurement, and Phenomena}, 18\penalty0 (3):\penalty0 1426--1430, 2000.
\newblock \doi{10.1116/1.591396}.

\bibitem[Xu et~al.(2016)Xu, Huang, Liu, Xu, Ma, and Chu]{xu2016effects}
HY~Xu, YH~Huang, S~Liu, KW~Xu, F~Ma, and Paul~K Chu.
\newblock Effects of annealing ambient on oxygen vacancies and phase transition temperature of \ch{VO2} thin films.
\newblock \emph{RSC advances}, 6\penalty0 (83):\penalty0 79383--79388, 2016.
\newblock \doi{10.1039/C6RA13189A}.

\bibitem[Popov et~al.(2001)Popov, Monge, Gonz{\'a}lez, Chen, and Kotomin]{popov2001dynamics}
AI~Popov, MA~Monge, R~Gonz{\'a}lez, Y~Chen, and EA~Kotomin.
\newblock Dynamics of f-center annihilation in thermochemically reduced \ch{MgO} single crystals.
\newblock \emph{Solid state communications}, 118\penalty0 (3):\penalty0 163--167, 2001.
\newblock \doi{10.1016/S0038-1098(01)00062-X}.

\bibitem[Kingma and Ba(2014)]{kingma2014adam}
Diederik~P Kingma and Jimmy Ba.
\newblock Adam: A method for stochastic optimization.
\newblock \emph{arXiv preprint arXiv:1412.6980}, 2014.

\end{thebibliography}






\clearpage

\renewcommand{\thefigure}{S\arabic{figure}}
\setcounter{figure}{0}
\renewcommand{\thetable}{S\arabic{table}}
\setcounter{table}{0}
\renewcommand{\theequation}{S\arabic{equation}}
\setcounter{equation}{0}

\section*{Supplemental Information}

\subsection*{Additional growth details}

Si substrates are cleaned prior to growth using a modified HF-last Radio Corporation of America (RCA) cleaning process consisting of a 10 minute soak in Standard Clean 1 (SC1) at 80 \celsius{}, 5 minutes of sonication in deionized water (DI), and a 30 second dip in hydrofluoric acid followed by a DI rinse \cite{kern2018cleaning}. Cleaned Si substrates are then outgassed in a dedicated vacuum chamber attached to the MBE at 300 \celsius{} for 30 minutes before introduction to the growth chamber. Upon loading into the growth chamber, substrates are heated until the Si(111) surface achieved a $7\times7$ reconstruction (between 785-835 \celsius{}) as observed with the \textit{in situ} RHEED system. Substrate temperature is measured using an optical pyrometer calibrated against the Al-Si eutectic point at 577 \celsius{} \cite{bertness2000smart}. Er and Ce are evaporated using resistively heated effusion cells (Riber HT-12) and ultra-high purity molecular oxygen (\ch{O2}) is provided via an MKS mass flow controller. Beam equivalent fluxes of the Ce and \ch{O2} are determined using a beam flux monitor to set a ratio of \ch{O2} to Ce of $\sim20$ during growth, with growth rates of 250-360 nm/hr.

\subsection*{Analyzing threading dislocations via XTEM}

\begin{figure}[ht]
  \centering
  \includegraphics[width=0.8\linewidth]{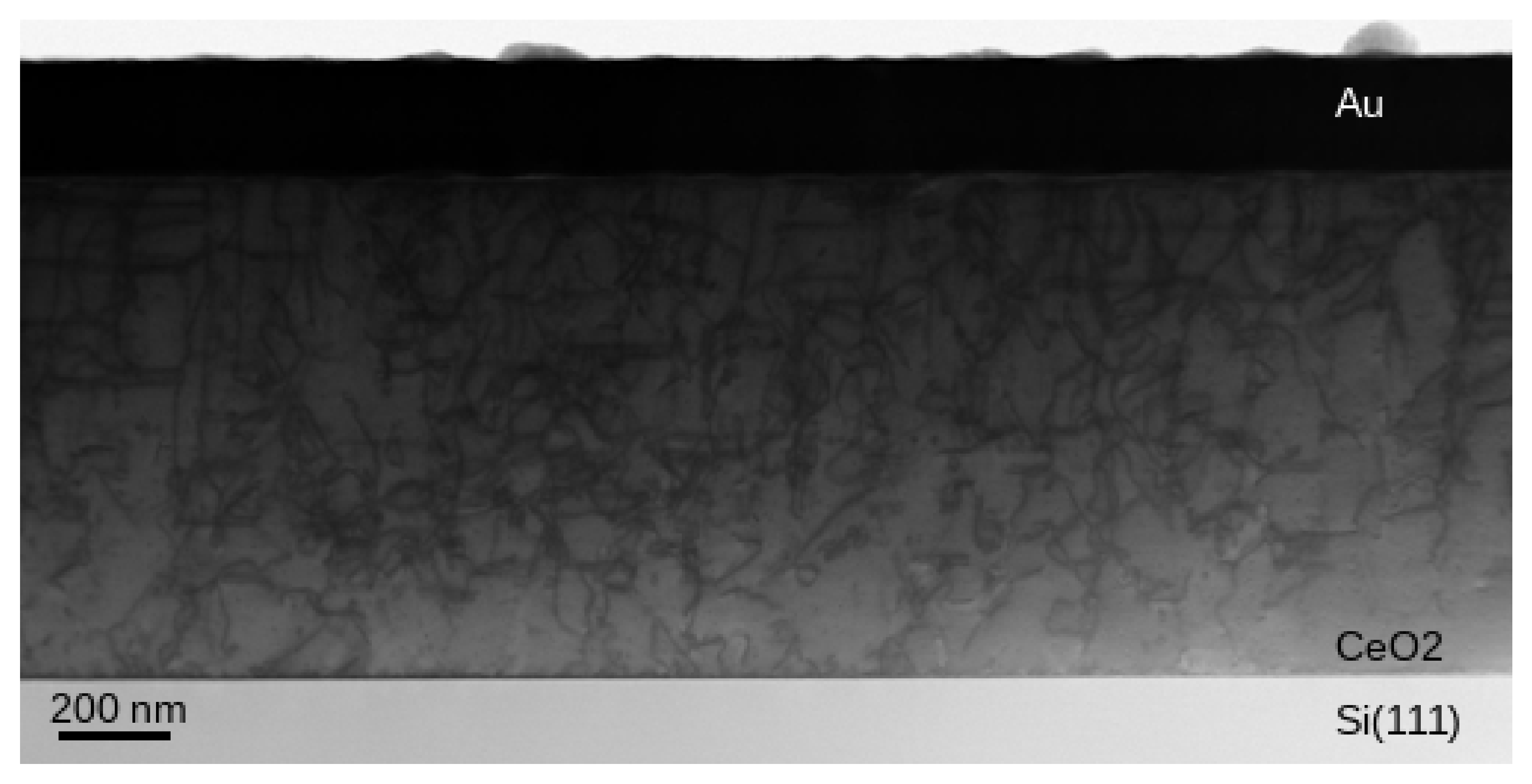}
  \caption{Low magnification bright field cross-sectional TEM of the full thickness of an as-grown \ch{Er:CeO2} film (940 nm thick, 3 ppm erbium) on silicon, showing the presence of numerous threading dislocations, with an areal density of approximately $3.3\times10^{9}$ dislocations per cm${^2}$.}
  \label{si:fig1}
\end{figure}

\subsection*{Additional details of doping series samples}

Figures~\ref{fig:fig3} and~\ref{fig:fig4} show results of varying the concentration of Er in \ch{CeO2} films.  The thicknesses of these films vary between samples, and so we have listed those thicknesses here.

\begin{table}[h]
    \centering
    \caption{Additional details on samples used in the doping series discussed in the main text.}
    \begin{tabular}{|l||l|l|l|l|l|l|l|}
    \hline
    Er Doping Level (ppm) & 2   & 3   & 14.5 & 15  & 49  & 88  & 132 \\ \hline
    Thickness (nm)     & 760 & 940 & 740  & 800 & 775 & 860 & 835 \\ \hline
    \end{tabular}
    \label{si:tb1}
\end{table}

\subsection*{Fitting the Inokuti-Hiroyama model}

In Figure~\ref{fig:fig4}(c) we use the Inokuti-Hiroyama model \cite{wolf1976progress, inokuti1965influence} to analyze the results of the optical excited state lifetime of the $Z_1$-$Y_1$ transition. The model considers an optically active emitter of intrinsic lifetime $T_1^{(0)}$ surrounded by defects that can quench the excitation. The energy transfer from the emitter to the defects may occur via dipole-dipole interaction or via higher order (dipole-quadrupole, etc.) interactions. The resulting radiative decay lifetime $T_1$ is given by the following two equations \cite{wolf1976progress}:

\begin{equation}
    \phi(t) = \phi_0 \exp \Big[-\frac{t}{T_1^{(0)}} - \Gamma\Big(1-\frac{3}{\nu}\Big) \frac{n}{n_0} \Big(\frac{t}{T_1^{(0)}}\Big)^{3/\nu} \Big]
    \label{ih}
\end{equation}

\begin{equation}
    T_1 = \int_{0}^{\infty} t \phi(t) dt \Big/ \int_{0}^{\infty} \phi(t) dt
    \label{ih-integrate}
\end{equation}

These equations yield the mean lifetime of the emitter $T_1$ as a function of the intrinsic lifetime of the emitter ($T_1^{(0)}$) surrounded by a concentration of quenching defects $n$ (\# per volume). Here the parameter $n_0$ represents a `critical concentration' of the quenching defects (\# per volume).  The type of interaction is governed by the parameter $\nu \in \{6, 8, 10\}$ for electric dipole-dipole, electric dipole-quadrupole, and electric quadrupole-quadrupole interactions respectively.  Finally, the excited state population as a function of time $\phi(t)$ is normalized by the parameter $\phi_0$ representing the the number of emitters.

To fit our lifetime data as a function of erbium doping level using the Inokuti-Hiroyama model, we fit for $T_1$ by using the intrinsic lifetime $T_1^{(0)}$ and critical concentration $n_0$ as free parameters.  The normalization constant $\phi_0$ is factored out in Equation~\ref{ih-integrate}.  Numerical integration is used to evaluate Equation~\ref{ih-integrate}.

Running the fit on the lifetime data with $\nu=6$ produces an intrinsic lifetime of $T_1^{(0)} = 3.42(3)$ ms as discussed in the main text, and also yields a critical concentration of $n_0 = 474(46)$ ppm, though the identity of the quenching defect is not discussed.  We choose $\nu=6$ due to the electric dipole-dipole interaction being the longest-range interaction, but fitting $\nu=8$ and $\nu=10$ yield fits of equivalent quality with the same intrinsic lifetime, the only change being that the critical concentrations are fitted to $219(21)$ ppm and $295(28)$ ppm respectively.

\subsection*{Full treatment of the annealing derivation and conclusions}

With annealing, the density of charged defects drops, resulting in narrowing inhomogeneous broadening and lengthening $T_1$ of the Er emitters.  Here we assume a kinetic equation with first order reaction rate for the annealing of the charged defects as a function of time: 

\begin{equation}
    \frac{dn}{dt} = -K\big (n-n_f)
    \label{rate_eqn}
\end{equation}

Here, $n$ is the density of charge defects, $K$ is the reaction rate constant, and $n_f$ is the defect concentration at thermal equilibrium. The thermally activated reaction rate constant $K$ is given by the Arrhenius relation $K \propto \big(\exp{-E_A/k_B T}\big)$ where $E_A$ is the activation energy for the defect to hop between lattice sites, $k_B$ is the Boltzmann constant, and $T$ is the sample temperature.  Far from equilibrium, assuming $n \gg n_f$, we have from Equation ~\ref{rate_eqn}:

\begin{equation}
    n(t) = n_i \exp \big(-Kt\big) \text{ , } K \propto \exp\big(-E_A/k_B T\big)
    \label{nvst}
\end{equation}

where the defect density drops to $n(t)$ after annealing for a duration $t$ from the initial ``grown-in" concentration $n_i$.  Similar dependencies to Equation ~\ref{nvst} have been seen in other works on similar polar oxides \cite{xu2016effects, popov2001dynamics} as well.

The power-law dependence of the inhomogeneous linewidth on concentration is established in Figure~\ref{fig:fig4}(b). As discussed, some ambiguity in the appropriate power-law remains. Thus, without loss of generality we may apply the generic model of $\Gamma_{\text{inh}} = a + b n_t^\alpha $ where $n_t$ is the density of grown-in defects (that impact the linewidth) remaining after dwelling for time $t$ during annealing, and $\alpha$ is the generic exponent. It is possible that these as-grown defects affect $\Gamma_{\text{inh}}$ through strain interactions, in which case $\alpha = 1$, and the power law dependence reverts to that in Equation~\ref{inh} with $b=0$. Combining this generic power law with Equation~\ref{nvst} we thus have:

\begin{equation}
    \Gamma_{\text{inh}} = a + b n_i^\alpha \exp \big(- \alpha K t\big)
\end{equation}

Since we do not know the intrinsic number of defects and the duration of annealing is held constant between samples, we absorb those parameters along with the generic exponent $\alpha$ into the fit parameters $b'$ and $c$ to obtain the following function to be used for fitting:

\begin{equation}
    \Gamma_{\text{inh}} = a + b' \exp \Big(-c \exp \big(-E_A / k_B T\big)\Big)
    \label{kinetic}
\end{equation}

where $a$, $b'$, $c$, and $E_A$ are fitting parameters. Fitting using an adaptive gradient descent optimizer \cite{kingma2014adam} yields the fitting parameters $a=6.21$ GHz, $b'=2.74$ GHz, $c=1094$, and the fitted activation energy $E_A = 0.663$ eV for the inhomogeneous linewidth, indicating consistency with our model for a reaction-limited annihilation of defects under annealing, with a reaction rate that is thermally activated, likely diffusion limited, and has an Arrhenius-like dependence on temperature.

The increase of the excited state lifetime with increased annealing temperature is modeled with the Inokuti-Hirayama approach as discussed before.  Again, we assume in our model that the population of the quenching defects surrounding the erbium emitters are annealed out -- for which a thermally activated kinetic model, same as Equation ~\ref{nvst} is considered. Thus, in Equation ~\ref{ih} and ~\ref{ih-integrate}, we take $n=n_i \exp \big(-Kt\big) \text{ , } K \propto \exp\big(-E_A/k_B T\big)$. This leads to the following equations for $T_1$ as a function of annealing time $t$:

\begin{equation}
    \phi(t') = \phi_0 \exp \Big[-\frac{t'}{T_1^{(0)}} - \Gamma\Big(1-\frac{3}{\nu}\Big) \Big(\frac{n_i}{n_0}\Big)\exp \big(-Kt\big) \Big(\frac{t'}{T_1^{(0)}}\Big)^{3/\nu} \Big],  K =c\exp\big(-E_A/k_B T\big)
    \label{ih_anneal}
\end{equation}

\begin{equation}
    T_1 = \int_{0}^{\infty} t' \phi(t') dt' \Big/ \int_{0}^{\infty} \phi(t') dt'
    \label{ih-integrate_anneal}
\end{equation}

Here $T_1^{(0)}$, $(n_i/n_0)$, $c$ and $E_A$ act as free parameters for fitting. Again, fitting using an adaptive gradient descent approach leads to a good fit (Figures~\ref{fig:fig5}(c)) with $T_1^{(0)}=5.72$ ms, $(n_i/n_0)=0.792$, $c=946$, and $E_A = 0.718$ eV. We have taken $\nu=6$ representing ED-ED interaction. This is in the same regime of $E_A$ arrived at for the inhomogeneous linewidth variation, and points to a common density of grown in defects that are being annihilated via thermally activated processes, thus leading to improvement of the optical characteristics of the Er emitters.

\end{document}